# Observation of Topological Corner State Arrays in Photonic Quasicrystals

*Aoqian Shi, Yiwei Peng, Jiapei Jiang, Yuchen Peng, Peng Peng, Jianzhi Chen, Hongsheng Chen, Shuangchun Wen, Xiao Lin, Fei Gao,\* and Jianjun Liu\**

A. Shi, J. Jiang, Y. Peng, P. Peng, J. Chen, S. Wen, J. Liu
Key Laboratory for Micro/Nano Optoelectronic Devices of Ministry of Education & Hunan Provincial Key Laboratory of Low-Dimensional Structural Physics and Devices, School of Physics and Electronics, Hunan University, Changsha 410082, China
E-mail: jianjun.liu@hnu.edu.cn (Jianjun Liu)

Y. Peng, H. Chen, X. Lin, F. Gao
Interdisciplinary Center for Quantum Information, State Key Laboratory of Extreme Photonics and Instrumentation, ZJU-Hangzhou Global Scientific and Technological Innovation Center, Zhejiang University, Hangzhou 310027, China
E-mail: gaofeizju@zju.edu.cn (Fei Gao)

Y. Peng, H. Chen, X. Lin, F. Gao
International Joint Innovation Center, The Electromagnetics Academy at Zhejiang University, Zhejiang University, Haining 314400, China

J. Liu
Greater Bay Area Institute for Innovation, Hunan University, Guangzhou 511300, China

Recently, the studies of topological corner states (TCSs) are extended from crystals to quasicrystals, which are referred to as higher-order topological quasicrystalline insulators (HOTQIs). However, the TCSs of complete quasi-periodic structure in photonic systems have yet to be demonstrated. Moreover, there is only one TCS in each corner region in higher-order topological insulators (HOTIs). Increasing the number of TCS is expected to increase application potential of TCSs. In this work, HOTQIs in photonic systems are experimentally observed. It is found that HOTQIs possess TCS arrays, and each TCS array contains several

TCSs. Furthermore, the universal theoretical framework of the multimer analysis method is improved, and the difference in the average charge density is proposed as a real-space topological index. These results will open up new ideas for investigating highly integrated, multi-region localized TCSs and are expected to provide new ways to explore topological phenomena and the applications of photonic quasicrystals.

## 1. Introduction

Higher-order topological insulators (HOTIs) based on spatial symmetries (such as rotational symmetry and mirror symmetry) are characterized by higher-order topological states with lower dimensions, such as one-dimensional (1D) topological hinge states in 3D systems and 0D topological corner states (TCSs) in 2D or 3D systems.[1–20] HOTIs have been theoretically and experimentally demonstrated in electronics,[1–6] photonics,[7–15] and phononics.[16–20] HOTIs provide an excellent platform for studying novel physical phenomena, such as bound states in the continuum[21,22] and topological defects (such as dislocation states and disclination states).[23–27] Furthermore, the combination of HOTIs with systems beyond linear band topology, including non-Hermitian, nonlinear and non-Abelian systems, may lead to unconventional phenomena.[28]

In addition to fundamental physics, HOTIs also play a unique role in applied physics. The 0D TCSs have been used in coupled cavity-waveguide,[29] topological photonic crystal fibers,[30] and topological lasers.[31] Since there is only one TCS in each corner region, the integration of TCSs (i.e., the number of TCSs in each corner region) is limited. Increasing the integration of TCSs is expected to increase the application potential of TCSs. Recently, some theoretical and experimental studies found that considering the special case where long-range couplings (LRCs) stronger than nearest-neighbor couplings (NNCs) can extend the HOTI of the $Z_2$ class to the $Z$ class, the number of TCSs in each corner region can be increased.[32–34] However, achieving large-number TCSs in each corner region remains a challenging task in the general case where NNCs are stronger than LRCs.

Different from crystals, 2D quasicrystals (quasi-periodic systems) exhibit rotational symmetry, self-similarity, and long-range order.[35–39] Higher-order topological quasicrystalline insulators (HOTQIs) based on quasicrystals exhibit unique features that are absent from crystals (such as TCSs protected by $C_5$, $C_8$, and $C_{12}$ symmetries),[40–47] and HOTQIs are expected to possess novel topological phenomena including highly integrated TCSs. Until now, the studies of HOTQIs have been almost based on electronic systems. Photonic systems have only recently been theoretically studied based on the 2D Thue–Morse structure.[47] Moreover, this structure is not a quasi-periodic lattice but is arranged in a square lattice. Therefore, the connection between higher-order topology (HOT) and the complete quasi-periodic structure in photonic and phononic systems has yet to be established. In addition, to the best of our knowledge, HOT in quasicrystal has only been experimentally observed in electrical circuits.[46] Although first-order and higher-order topological states have been experimentally verified in fractals (another noncrystalline system besides quasicrystals),[18,19,48] fractals are not equivalent to quasicrystals. Fractals rely on self-similarity, possess fractal dimensions, and lack conventional bulk regions,[48] resulting in physical properties that differ from those of quasicrystals. The experimentally realized HOTQI in electrical circuits needs to be introduced negative couplings,[46] which makes it inconvenient to extend to photonic and phononic systems. Acknowledging the extensive potential applications of photonic and phononic systems,[28] the pursuit of suitable methods to realize HOTQIs within these systems is of great significance.

In this work, Stampfli-type and Stampfli-ring-type HOTQIs in photonic systems are theoretically analyzed and experimentally verified. Two HOTQIs contain three corner regions (i.e., corner-I, corner-II and corner-III regions). It is found that the eight (twelve) TCSs in each corner-II (corner-III) region form an array and exhibit a polyline-shaped (ring-shaped) distribution. The universal theoretical framework of the multimer analysis method (MAM) is improved, and the difference in the average charge density (*ACD*) is proposed as a real-space topological index, which characterizes the HOT.

## 2. Results and Discussion

### 2.1. Tight-binding models and MAM

Using an iterative subdivision method, with squares and regular triangles as primitive tiles, a Stampfli-type quasicrystal tiling can be constructed (see Section A in Supporting Information). The vertex of each primitive tile (i.e., the lattice site of the quasicrystal) is used as a basic cell. Since the Stampfli-type quasicrystal satisfies the $C_6$ symmetry, considering the interaction between adjacent basic cells, each basic cell contains six subsites. Subsites can set scatterers corresponding to electronic, photonic, and phononic systems. By removing seven basic cells in the center region of a Stampfli-type quasicrystal, a Stampfli-ring-type quasicrystal can be constructed. HOTs in Stampfli-type and Stampfli-ring-type quasicrystals are investigated, as shown in **Figure 1**.

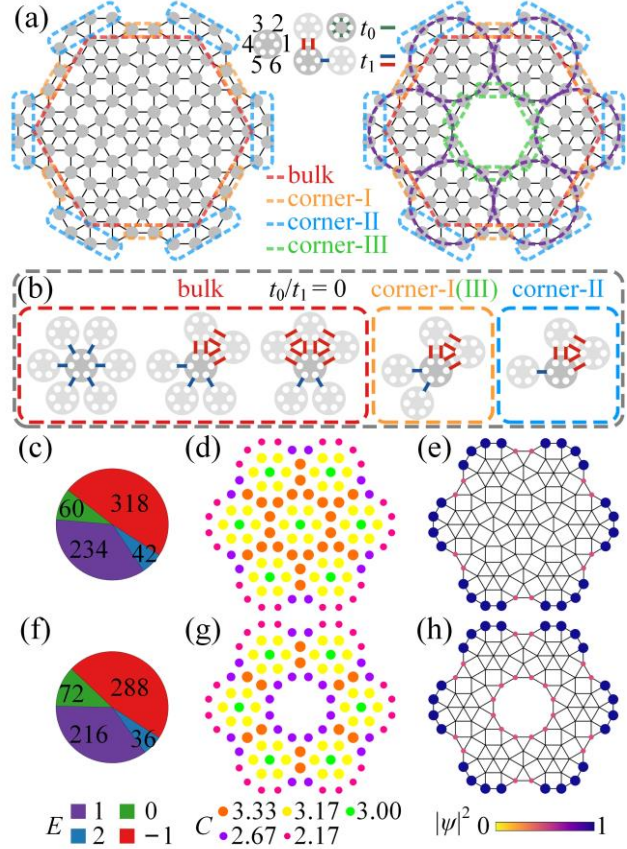

**Figure 1.** Tight-binding models and HOT of Stampfli-type and Stampfli-ring-type quasicrystals based on MAM. a) Tight-binding models, which contain 109 and 102 basic cells, respectively. The dashed boxes indicate the bulk (red) and corner regions (yellow, blue, and green). Due to the properties of quasicrystals, the corner-I region corresponds to the edge region of previous HOTIs, and the corner-II region is arranged in a polyline. The Stampfli-ring-type quasicrystal consists of six equivalent units (purple dashed circles). The insets show the serial number of subsites in the basic cell as well as the schematic of the NNCs within the basic cell ($t_0$) and between adjacent basic cells ($t_1$). b) Multimer types in the two quasicrystals. The gray circle represents the basic cell in the bulk and corner regions, and the light gray circles represent the adjacent basic cells. Energy spectrum of the eigenvalues, charge density, and normalized probability distribution of the wave function with the parameters $t_0 = 0$ and $t_1 = 1$ for c)–e) the Stampfli-type HOTQI and f)–h) the Stampfli-ring-type HOTQI, respectively.

In a finite-size system, the tight-binding Hamiltonian of the two quasicrystals shown in Figure 1a can be expressed as

$$H_{\text{Stampfli}} = \sum_i c_i^\dagger t_0 h_0 c_i + \sum_{\langle i,j \rangle} c_i^\dagger t_1 h_1 c_j, \tag{1}$$

where $c_i^\dagger$ represents the electron creation operator of the basic cell $i$, $c_i$ ($c_j$) represents the electron annihilation operator of the basic cell $i$ ($j$), and $h_0$ ($h_1$) is determined by the NNCs within (between) basic cells. The two quasicrystals satisfy the long-range $C_{12}$ symmetry dominated by the $C_6$ symmetry (see Section B in Supporting Information). The tight-binding Hamiltonian with LRCs is further discussed in Section C in Supporting Information.

When $t_0$ ($t_1$) = 0 and $t_1$ ($t_0$) ≠ 0, the tight-binding model behaves as a combination of multimers (see Section D in Supporting Information). According to the symmetry of the two quasicrystals, there are hexamers, trimers, dimers, and monomers. The specific multimer forms in the two quasicrystals for the case where $t_0 = 0$ and $t_1 = 1$ are shown in Figure 1b. The two quasicrystals have one monomer (two monomers) in each basic cell of the corner-I (corner-II) region, while there are no monomers in the bulk regions. For the Stampfli-ring-type quasicrystal, the multimer type in the corner-I region is consistent with that in the corner-III region.

The energy spectrum ($E$) of the eigenvalues is obtained by solving the Hamiltonian $H_{Stampfli}$. The two quasicrystals have 60 and 72 zero-energy states, respectively, as shown in Figure 1c and f, respectively. The number of zero-energy states in the energy spectrum is equal to the number of monomers in the tight-binding model (see Section E in Supporting Information). It can be seen from Figure 1a and b, for Stampfli-type quasicrystal, a total of 12 (48) monomers are contained in six corner-I (corner-II) regions, and 60 monomers correspond to 60 zero-energy states. For Stampfli-ring-type quasicrystal, although the multimer type is consistent within corner-I and corner-III regions, the effect of monomers does not overlap but rather accumulates. Therefore, an additional 12 monomers from six corner-III regions are added on the basis of the Stampfli-type quasicrystal, resulting in an additional 12 zero-energy states in Stampfli-ring-type quasicrystal. Filling the band with $E < 0$, the charge density ($C$) of the two quasicrystals is shown in Figure 1d and g, respectively. The corner charge density is different from the bulk charge density; this corner anomaly is a signature of the HOT.[4,5,43,44] From the probability distribution of the wave function of the two quasicrystals shown in Figure 1e and h, respectively, it can be found that there are TCSs protected by the symmetry of the two quasicrystals in the corner regions, which is a sign that the two quasicrystals possess HOT, indicating the realization of two HOTQIs. Interestingly, different from conventional TCSs, there are TCS arrays with a polyline-shaped distribution in the two quasicrystals, which contain eight TCSs in the four basic cells of each corner-II region. The Stampfli-ring-type HOTQI also possesses a TCS array with a ring-shaped distribution (12 TCSs in the corner-III regions). Furthermore, TCS arrays persist when $t_0$ is assigned a non-zero value (see Section F in Supporting Information).

## 2.2. Characterization of HOT

The corner anomaly was used as a real-space topological index in previous studies of HOTIs.[4,5] Furthermore, a recent study has utilized the local density of states of photonic crystals to characterize the non-trivial bulk topology of the system.[49] In this work, the relationship between the charge density and the multimer type is demonstrated by improving the theoretical framework of the MAM (see Section D and E in Supporting Information). For the band with $E < 0$, the *ACD* of the subsites corresponding to the non-monomers in the structure is

$$ACD_0 = \frac{1}{S_0} \sum_{s=1}^{S_0} \sum_{\alpha=1}^{N} |\psi_s^\alpha|^2, \tag{2}$$

and the *ACD* of the subsites corresponding to the monomers in the corner-I (corner-II) region is

$$ACD_{1(2)} = \frac{1}{S_{1(2)}} \sum_{s=1}^{S_{1(2)}} \sum_{\alpha=1}^{N} |\psi_s^\alpha|^2, \tag{3}$$

where $\psi_s^\alpha$ is the wave function amplitude at the subsite $s$ for the eigenstate $\alpha$, $N$ is the total number of eigenstates below the band gap, and $S_{1(2)}$ is the total number of subsites corresponding to the monomers in the corner-I (corner-II) region. The $S_{1(2)}$ of the Stampfli-type and Stampfli-ring-type quasicrystals are 12 (48) and 24 (48), respectively. Since the multimer types in the corner-I and corner-III regions in the Stampfli-ring-type HOTQI are the same, $S_1$ is the sum of the number of monomers in the two regions. $S_0$ is the total number of subsites corresponding to the non-monomers. The $S_0$ of the two quasicrystals are 594 and 540, respectively.

When the system is in the trivial phase, the *ACD* of any subsite in the system is very similar, and it is exactly the same at $t_1/t_0 \approx 0$. When the system is in the topological phase, the electrons below the band gap mainly fill the non-monomers, and the *ACD* of the subsite corresponding to the monomer is almost 0. When $t_0/t_1 \approx 0$, the *ACD* is actually 0. This is because only zero-energy ($E = 0$) electrons can fill the monomers (see Section D and E in Supporting Information). $E = 0$ is in the band gap; thus, the non-zero-energy electrons below the band gap cannot fill the monomer, causing its *ACD* to be 0. The difference between $ACD_0$ and $ACD_{1(2)}$ can be used as a real-space topological index to effectively characterize the topology of the system, which can be expressed as

$$\Delta ACD = \frac{ACD_0 - ACD_{1(2)}}{\max(ACD_0) - \min(ACD_{1(2)})}, \tag{4}$$

where $\Delta ACD$ is normalized, $\max(ACD_0) = N/S_0$, and $\min(ACD_{1(2)}) = 0$. The $\Delta ACD$ is 0 (non-0) for the trivial (topological) phase, and the gapless metallic region with $\Delta ACD$ close to 0 is not included.

The variation in $\Delta ACD$ with $t_0$ and $t_1$ is shown in **Figure 2**.

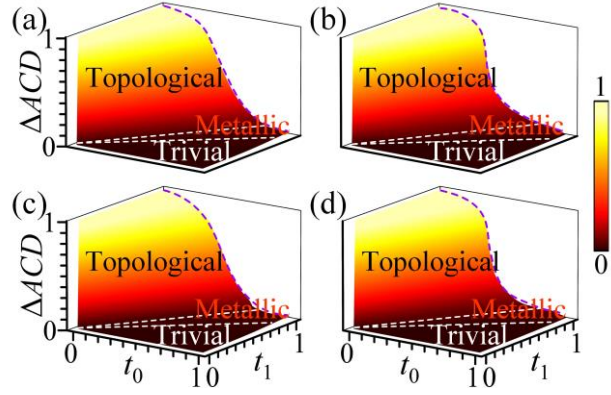

**Figure 2.** Variation in $\Delta ACD$ with $t_0$ and $t_1$. $\Delta ACD$ for $ACD_0$ and $ACD_1$: a) Stampfli-type quasicrystal, and c) Stampfli-ring-type quasicrystal. $\Delta ACD$ for $ACD_0$ and $ACD_2$: b) Stampfli-type quasicrystal, and d) Stampfli-ring-type quasicrystal.

For the Stampfli-type quasicrystal, by comparing Figure 2a and b, considering that $t_1 = 1$ and $t_0$ decreases (i.e., $t_1/t_0$ increases), it can be seen that the variation in the two $\Delta ACD$s (purple dashed lines) is not the same. The growth of $\Delta ACD$ in Figure 2a is relatively gradual, while $\Delta ACD$ in Figure 2b increases significantly when $t_1/t_0$ is large. This indicates that with the increase in $t_1/t_0$, the TCSs of the corner-I region appear earlier than those of the corner-II region, which is consistent with the fact that the TCSs appear after the topological edge state reported in previous HOTIs.[6,7] The $\Delta ACD$ of the Stampfli-ring-type quasicrystal is consistent with that of the Stampfli-type quasicrystal, as shown in Figure 2c and d. By further combining the energy spectrum, charge density and multimer types, the number of TCSs contained in the TCS arrays can also be characterized (see Section G in Supporting Information). The MAM and $\Delta ACD$ provide an effective way to characterize the HOT of various systems. The HOTIs and HOTQIs with different symmetries are further discussed in Section H and I in Supporting Information. Moreover, the HOTQI realized in this work is not significantly influenced by the shape of quasicrystal. To verify its universality, the tight-binding Hamiltonian is extended to Ammann-Beenker and Equithirds quasicrystals selected randomly[50] and two HOTQIs are realized (see Section J in Supporting Information).

### 2.3. Stampfli-type photonic HOTQI

Based on the theoretical model of the electronic system, HOTQIs in photonic system (i.e., photonic HOTQIs) are simulated and experimentally demonstrated. The Stampfli-type photonic HOTQI is shown in **Figure 3**.

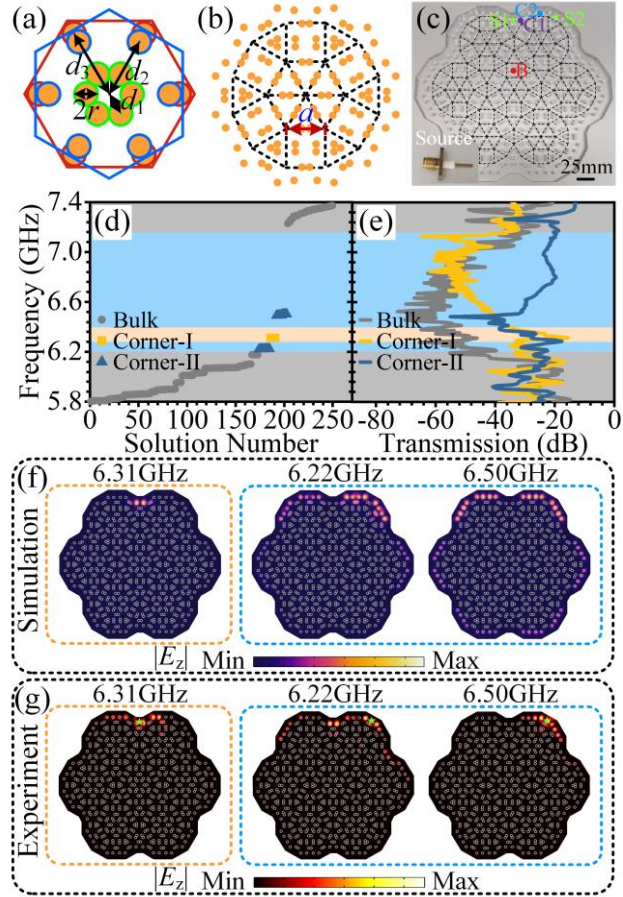

**Figure 3.** Stampfli-type photonic HOTQI. a) Schematic of the basic cell, b) schematic of 19 basic cells, and c) photograph of the sample. The scatterer material is a zirconia ceramic ($\varepsilon_{r1}$ = 26), the radius is $r$ = 2.5 mm, and the height is 10 mm. The distance between two adjacent basic cells is $a$ = 25 mm, and the distance $d$ ($d$ = $d_1$, $d_2$, $d_3$) between the scatterer and the center of the basic cell depends on the multimer type, that is, $d_1$ = $2r$ (hexamer), $d_2$ = $a/2-r$ (dimer), and $d_3$ = $(a-2r)/\sqrt{3}$ (trimer). The black dashed frame shows the quasicrystal structure. d) Simulated eigenfrequency distribution. e) Measured transmission spectrum. The excitation sources are located at S1 and S2, and the probes are located at B, C1, and C2; these points are marked in panel c). TCSs in the corner-I and corner-II regions derived from f) eigenstate simulations and g) experimental measurements. The green pentagram marks the position of the excitation source.

As shown in Figure 3a–c, based on the tight-binding model of the Stampfli-type quasicrystal in Figure 1a, six dielectric scatterers are placed at six subsites in each basic cell. The distance between the scatterer and the center of the basic cell is adjusted to construct the different multimers (see Section K in Supporting Information). Different from previous HOTIs that used crystals as the trivial cladding in simulations and experiments, this work uses a metal (perfect electric conductor, PEC) as the trivial cladding to make the structure more compact. From Figure 3d and e, it can be seen that there are TCSs corresponding to the monomers of the two corner regions in the band gap range of $f \in$ [6.20 GHz, 7.16 GHz] (see Section L in Supporting Information). The electric field distribution is shown in Figure 3f and g. The simulation and experimental results are in good agreement with each other, and the TCS array (including eight TCSs) with a polyline-shaped distribution in the corner-II region verifies the correctness of the theoretical model.

## 2.4. Stampfli-ring-type photonic HOTQI

The Stampfli-ring-type photonic HOTQI is investigated, as shown in **Figure 4**.

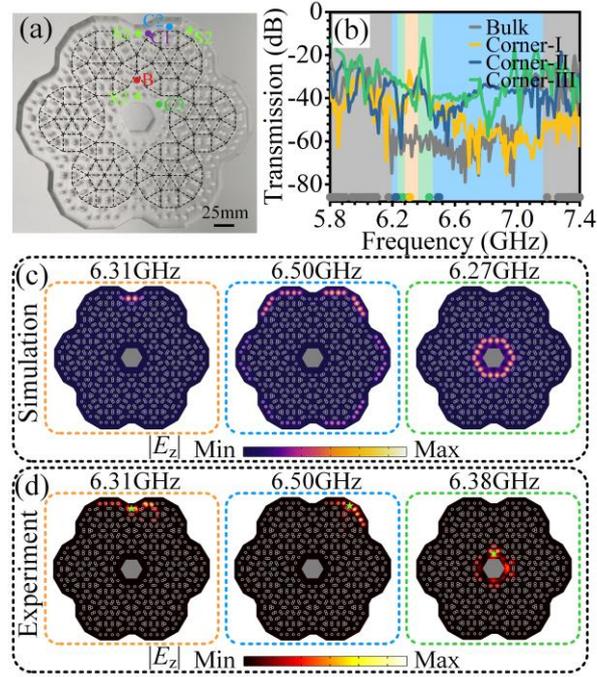

**Figure 4.** Stampfli-ring-type photonic HOTQI. a) Photograph of the sample. The seven basic units in the central region are removed, and they are replaced with a hexagonal metal prism. The black dashed frame shows the quasicrystal structure. b) Measured transmission spectrum. The excitation sources are located at S1, S2, and S3, and the probes are located at B, C1, C2, and C3; these points are marked in panel a). The gray, yellow, blue, and green dots on the horizontal axis correspond to the simulated eigenfrequencies of the bulk and corner states. TCSs in the corner-I, corner-II, and corner-III regions derived from c) eigenstate simulations and d) experimental measurements. The green pentagram marks the position of the excitation source.

From Figure 4b, it can be found that there are TCSs corresponding to the monomers of the three corner regions in the band gap range of $f \in$ [6.20 GHz, 7.16 GHz]. The electric field distribution is shown in Figure 4c and d. In the corner-II region, there is a TCS array (including eight TCSs) with a polyline-shaped distribution, and in the corner-III region, there is a TCS array (including 12 TCSs) with a ring-shaped distribution, which verifies the correctness of the theoretical model. Compared with the Stampfli-type photonic HOTQI, although seven basic cells are removed from the bulk region, the TCSs in the corner-I and corner-II regions remain unchanged.

## 2.5. Robustness of photonic HOTQI

Three types of defects are introduced into the quasicrystal to investigate the robustness of photonic HOTQI, as shown in **Figure 5**.

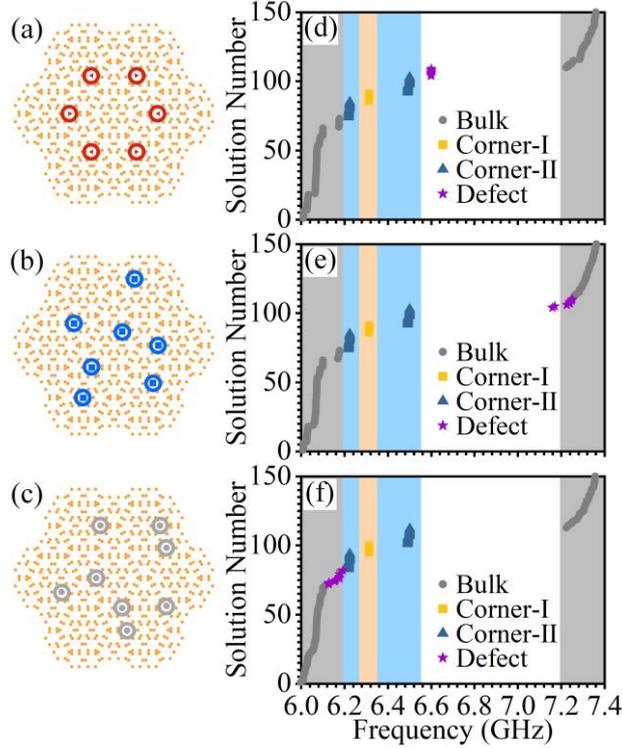

**Figure 5.** Robustness of photonic HOTQI. a)–c) Schematic of defects in Stampfli-type photonic HOTQI. Types of defects: a) six triangular scatterers, the side lengths are all $l_d = 3r$ and the materials are all germanium $\varepsilon_{r1} = 16$. b) seven square scatterers, the side lengths are all $l_d = 4r$, and the materials are all silica glass $\varepsilon_{r1} = 2.1$. c) eight circular scatterers, the radii are all $r_d = 2r$, and the materials are all PEC. The red, blue, and gray circles correspond to the defect positions, and the defects are all located in the interval of the original scatterers. d)–f) Eigenfrequency distributions corresponding to the three types of defects on the left.

In Figure 5a, the angles of the connecting lines between six defects and the center of the structure with respect to the horizontal direction are 0, $\pi/3$, $2\pi/3$, $\pi$, $4\pi/3$ and $5\pi/3$, respectively, and the quasicrystal keeps $C_6$ symmetry unchanged. In Figure 5b and c, the rotation symmetry of the quasicrystal is broken by introducing multiple randomly distributed defects. The eigenfrequency distributions of the quasicrystal under three conditions calculated by the finite element method are shown in Figure 5d–f. Although the introduction of defects produces local defect states in the band gap, and their frequencies are between the frequencies of the bulk states and the TCSs, the TCSs maintain excellent stability when defects are introduced, reflecting the robustness of photonic HOTQI.

In addition to photonic HOTQIs, phononic HOTQIs based on the MAM are theoretically demonstrated, and TCS arrays are realized (see Section M in Supporting Information). Photonic and phononic HOTQIs have an extraordinary application potential for use in several fields, such as high-efficiency lossless waveguides,[51,52] lasers,[53–55] and audio lasing,[56] due to their highly integrated, multi-region localized TCS arrays.

## 3. Conclusion

In this work, TCS arrays with polyline-shaped and ring-shaped distributions in Stampfli-type and Stampfli-ring-type photonic HOTQIs were demonstrated through the consistent results of theory, simulations, and experiments. The universal theoretical framework of the MAM was improved, and a real-space topological index was proposed. The experimentally realized photonic HOTQIs and theoretically realized phononic HOTQIs in this work broaden the classification of HOTIs, providing a fresh perspective for designing photonic and phononic devices with stronger localization and higher integration degree.

## 4. Experimental Section

*Numerical Simulations:* The finite element method is used for numerical simulation and the 3D models are constructed in simulations to better correspond to the experiments. For the eigenstate simulations in Figure 3f and 4c, the PEC boundaries are used in all directions.

*Experimental Measurements*: The experimental setups are mainly composed of two parts: the 2D near-field scanning platform (LINBOU, NFS02 Desktop Version 2D) and the vector network analyzer (Ceyear, 3672B). The detailed experimental setups and samples are presented in Section N in Supporting Information. By covering the top and bottom of the photonic HOTQI sample with aluminum plates, electric fields point in the out-of-plane $z$ direction and magnetic fields parallel to the $x$–$y$ plane to form a quasi-two-dimensional structure (TM mode). The metal cladding is used around the sample to equivalent the PEC boundary. The lower aluminum plate is placed above the stepper motor, which drives the sample platform to move with a step of 2.5 mm. There is an air layer of about 4 mm between the sample and the upper aluminum plate to facilitate the movement of the sample platform. The antenna (excitation source) is mounted on the lower aluminum plate through the drilled hole and located near the corner region of the photonic HOTQI. The probe antenna is fixed at the center of the upper aluminum plate. As the stepper motor drives the sample to move horizontally, the electric field $E_z$ can be measured.

**Supporting Information**

Supporting Information is available from the Wiley Online Library or from the author.

**Acknowledgements**

Aoqian Shi and Yiwei Peng contributed equally to this work. The authors thank Profs. Baile Zhang from Nanyang Technological University, Huaqing Huang from Peking University, Xiang Ni from Central South University, and doctoral student Xiaokang Dai from Hunan University for helpful discussions. Associate Prof. Jianjun Liu acknowledges the National Natural Science Foundation of China (Grants No. 61405058 and No. 62075059), the Natural Science Foundation of Hunan Province (Grants No. 2017JJ2048 and No. 2020JJ4161), and the Scientific Research Foundation of Hunan Provincial Education Department (Grant No. 21A0013). Profs. Fei Gao and Xiao Lin acknowledge the support partly from the National Natural Science Fund for Excellent Young Scientists Fund Program (Overseas) of China, the Key Research and Development Program of the Ministry of Science and Technology (Grants No. 2022YFA1404704, No. 2022YFA1405200, and No. 2022YFA1404902), the National Natural Science Foundation of China (Grants No. 61975176 and No. 62175212), the Key Research and Development Program of Zhejiang Province (Grant No. 2022C01036), Zhejiang Provincial Natural Science Fund Key Project (Grant No. Z23F050009), and the Fundamental Research Funds for the Central Universities (Grant No. 2021FZZX001-19).

**Conflict of Interest**

The authors declare no conflict of interest.

**Data Availability Statement**

The data that support the findings of this study are available from the corresponding author upon reasonable request.